\documentclass[aps,prb,twocolumn,showpacs,superscriptaddress,floatfix,citeautoscript]{revtex4}
\usepackage{times}
\usepackage{graphicx}
\usepackage{dcolumn}
\usepackage{bm}

\begin{document}

\title{
Incipient Wigner Localization in Circular Quantum Dots
}
\author{Amit Ghosal}
\affiliation{Department of Physics, Duke University,
Durham, North Carolina 27708-0305}
\affiliation{Physics Department, University of California Los Angeles, Los
Angeles, California 90095-1547}

\author{A.~D.~G\"u\c{c}l\"u}
\affiliation{Department of Physics, Duke University,
Durham, North Carolina 27708-0305}
\affiliation{Theory Center and Laboratory of Atomic and Solid State Physics, Cornell University,
Ithaca, New York 14853}

\author{C.~J.~Umrigar}
\affiliation{Theory Center and Laboratory of Atomic and Solid State Physics, Cornell University,
Ithaca, New York 14853}

\author{Denis Ullmo}
\affiliation{Department of Physics, Duke University,
Durham, North Carolina 27708-0305}
\affiliation{CNRS; Universit\'e Paris-Sud; LPTMS UMR 8626, 91405 Orsay Cedex, France}

\author{Harold~U.~Baranger}
\affiliation{Department of Physics, Duke University,
Durham, North Carolina 27708-0305}

\begin{abstract}
We study the development of electron-electron correlations in circular quantum dots as the density is decreased. We consider a wide range of both electron number, $N \!\le\! 20$, and electron gas parameter, $r_s \!\lesssim\! 18$, using the diffusion quantum Monte Carlo technique. Features associated with correlation appear to develop very differently in quantum dots than in bulk. The main reason is that translational symmetry is necessarily broken in a dot, leading to density modulation and inhomogeneity. Electron-electron interactions act to enhance this modulation ultimately leading to localization. This process appears to be completely smooth and occurs over a wide range of density. Thus there is a broad regime of ``incipient'' Wigner crystallization in these quantum dots. Our specific conclusions are:
(i)\,The density develops sharp rings while the pair density shows both radial and angular inhomogeneity.
(ii)\,The spin of the ground state is consistent with Hund's (first) rule throughout our entire range of $r_s$ for all $4 \!\le N \!\le\! 20$.
(iii)\,The addition energy curve first becomes smoother as interactions strengthen -- the mesoscopic fluctuations are damped by correlation -- and then starts to show features characteristic of the classical addition energy.
(iv)\,Localization effects are stronger for a smaller number of electrons.
(v)\,Finally, the gap to certain spin excitations becomes small at the strong interaction (large $r_s$) side of our regime.
\end{abstract}

\pacs{73.23.Hk, 73.63.Kv, 02.70.Ss}

\date{February 26, 2007}

\maketitle

\section{Introduction}

One of the fundamental quests in condensed matter physics research is to understand the effects of strong correlation between a system's constituent particles. A simplified system of particular interest is the ``electron gas" \cite{ElectronLiqBook}, in which conduction electrons interact pair-wise via Coulomb forces while the effect of atoms is ignored. It is well known that the electron gas has a Fermi liquid ground state with extended wave functions in the limit of high electron density, while when the density is decreased, thereby increasing the interaction strength, electrons become localized in space and order themselves in a ``Wigner crystal" phase \cite{ElectronLiqBook}. The interaction strength is often parametrized by the gas parameter $r_s \!=\! (c_d /a_B^*) (1/n)^{1/d}$ where $n$ is the electron density, $d$ is the spatial dimension, $a_B^*$ is the effective Bohr radius,
and $c_d$ is a dimension-dependent constant \cite{def_cd}.
In two dimensions, $r_s = 1/a_B^* (\pi n)^{1/2}$.
The physics at intermediate $r_s$ continues to offer puzzles, both from theory and experiments. There has been numerical evidence in bulk two \cite{Tanatar89,KwonCeperley93,Attaccalite02} and three \cite{ZongCeperley02,DrummondNeeds04} dimensional (2D and 3D) systems that a single transition takes place at $r_s^{\rm c,2D} \!\approx\! 30$-$35$ and $r_s^{\rm c,3D} \!\sim\! 100$. However, recent work in 2D has predicted more complex phases and associated transitions or crossovers around these critical values \cite{BarciOxman03,Spivak03,SpivakKivelson04,ChakravartyKNV99,JameiKivSpiv05,FalWaintal05,Waintal06}. While the experimental evidence is largely inconclusive \cite{2dMIRMP01,KravSarachik04}, and in particular the way in which the transition occurs isn't known experimentally, the problem has drawn a great deal of attention due to the hope of uncovering fundamental aspects of correlation effects.

Over the past decade or so, small confined systems, such as quantum dots (QD), have become very popular for experimental study \cite{MesoTran97,HeissQdotBook}. Beyond their possible relevance for nanotechnology, they are highly tunable in experiments and introduce level quantization and quantum interference in a controlled way. In a finite system, there can not, of course, be a true phase transition, but a cross-over between weakly and strongly correlated regimes is still expected. There are several other fundamental differences between quantum dots and bulk systems: (a)\,Broken translational symmetry in a QD reduces the ability of the electrons to delocalize. As a result, a Wigner-type cross-over is expected for a smaller value of $r_s$. (b)\,Mesoscopic fluctuations, inherent in any confined system \cite{MesoTran97,MesoHouches}, lead to a rich interplay with the correlation effects. These two added features make strong correlation physics particularly interesting in a QD. As clean 2D bulk samples with large $r_s$ are regularly fabricated these days in semiconductor heterostructures \cite{lowdens2DEG}, it seems to be just a matter of time before these systems are patterned into a QD, thus providing an excellent probe of correlation effects.

Circularly symmetric quantum dots have been a focal point of theoretical attention for several years \cite{ReimannMannRMP02}. Early calculations using density functional theory (DFT) within the local spin density approximation showed a spin density wave (SDW) signature for $r_s$ as small as $2$ \cite{KoskReimann97,BorghReimann05}. Unrestricted Hartree-Fock (UHF) \cite{YannLand99} yielded both SDW and charge density wave (CDW) features for $r_s \!\sim\! 1$. These were initially identified as signa\-tures of strong correlations related to the analog of a WC in a finite system, often called a ``Wigner molecule''. However, SDW and CDW both break the fundamental rotational symmetry of the 2D circular dots.
Indeed, later calculations confirmed that these effects are largely artifacts of the approxi\-mations used \cite{HiroseWin99,ReimannManninen00,HarjuNieminen02,ReuschGrabert03,PuenteNazm04,SzafPeeters04,BorghReimann05}. Projection methods were then used to restore symmetries as a second stage of the UHF calculations \cite{MikhailovaZieg02,YannLand03,YannLand04,YannLand06a,YannLand06b}. DFT has been pushed toward the large $r_s$ limit through the average spin density approximation \cite{GattobigioTosi05}. However, the validity of all these methods -- those that reduce an interacting problem to an effective noninteracting one -- remain questionable, particularly for large $r_s$.

More demanding computational techniques, which treat the correlations in an exact fashion, have also been applied to this problem. For example, exact diagonalization (ED) is accurate for small QDs \cite{ReimannManninen00,SzafPeeters04,Rontani06}, but becomes exponentially intractable for dots with more than 6 electrons ($N \!\geq\! 6$) and $r_s \!\geq\! 4$. Path Integral Monte Carlo (PIMC) has also been applied to large $r_s$ circular dots \cite{Egger99,Harting00,Filinov01,ReuschEgger03,WeissEgger05}. One study~\cite{Egger99} found a crossover from Fermi liquid to Wigner molecule behavior at $r_s \!\approx\! 4$, a value significantly smaller than the bulk $r_s^{\rm c,2D}$. Another \cite{Filinov01}, using different criteria for the transition, found a two-stage transition for $r_s$ larger than $r_s^{\rm c,2D}$. Although PIMC treats interactions accurately, it has its own systematic and statistical problems; for instance, it generates a thermal average of states with different $L$ and $S$ quantum numbers, preserving only $S_z$ symmetry. Since the energy for low-lying spin excitations becomes very small in the low density limit, the constraint on the temperature in order to access the ground state becomes extremely stringent. In the absence of a coherent picture, the role of correlations in QD remains an open problem.

The most accurate method for treating the ground state of strongly interacting quantum dots, in our opinion, is variational Monte Carlo (VMC) followed by diffusion Monte Carlo (DMC). This has been carried out in a number of cases for high to medium density quantum dots \cite{PederivaUmrigar00,CollettiUmrigar02,Pederiva02,GucluGuo03,GhosalUJUB05}. We discuss this method in more detail below. Briefly, while there is a systematic ``fixed-node error'', it is considerably smaller than the systematic and statistical errors of PIMC.

In this paper, we present a systematic study of circular parabolic QDs over a wide range of interaction strengths using an accurate VMC and DMC technique. We demonstrate how strong interactions bring out the interesting aspects of the Wigner physics in QDs; a short report on some aspects appeared in Ref.\,\onlinecite{GhosalNP06}. Our main results are:\\
(i) The development of inhomogeneities as the interaction strength increases is completely smooth, with no discernible special value below $r_s \!\approx\! 18$ for all $N \!<\! 20$. \\
(ii) The density develops sharp rings but remains circularly symmetric; the pair densities show both radial and angular inhomogeneity.\\
(iii) The spin of the ground state follows Hund's (first) rule throughout our entire range of $r_s$ for all $4 \!\le\! N \!\le\! 20$: it is the maximum consistent with the shell structure, in contrast to several previous claims \cite{Egger99,HarjuNieminen02,BorghReimann05,WeissEgger05,GattobigioTosi05}. We do find many violations of Hund's second rule -- the value of the orbital angular momentum is not necessarily a maximum -- but a modified second rule holds for small $r_s$. \\
(iv) The addition energy curve first becomes smoother as interactions strengthen -- the mesoscopic fluctuations are damped by correlation -- and then starts to show features characteristic of the classical addition energy, indicating incipient Wigner crystallization. \\
(v) Localization effects are stronger for a smaller number of electrons. \\
(vi) Finally, the gap to certain spin excitations becomes small at the strong interaction side of our regime.

The organization of the paper is as follows. In Section II, we describe the model and parameters for the quantum dots we study. Section III discusses our technical tools, VMC and DMC. We present our results in Section IV. We focus our attention first on the density and pair-densities: it is these two quantities that encode rich information on the correlation-induced inhomogeneities. We then discuss the energy of the ground and excited states. The interesting issue of spin-corre\-lation in a QD in the large $r_s$ limit is addressed at the end of the Section. Finally, we present our conclusions in Section V.

\section{Model and Parameters}\label{sec:model}

We consider quantum dots with $N$ electrons confined in a circularly symmetric harmonic potential, $V_{\rm con}({\bf r}) \!=\! m^* \omega^2 r^2/2$, using the Hamiltonian
\begin{equation}
{\cal H} = \sum_{i=1}^N \left(-\frac{\hbar^2}{2m^*} \nabla_i^2 +
V_{\rm con}({\bf r_i}) \right) + \sum_{i<j}^N \frac{e^2}{ 
\epsilon} \frac{1}{|{\bf r}_i - {\bf r}_j|}
\label{eq:Ham}
\end{equation}
where $m^*$ is the effective mass of the electrons and $\epsilon$ is the dielectric constant of the medium. We consider 2D systems, so that $r^2 = x^2 + y^2$. This is because experimental dots made by patterning GaAs/AlGaAs heterostructures have very strong confinement in the $z$-direction so that they are essentially two dimensional. The last term in the Hamiltonian is the pairwise Coulomb repulsion between electrons. The strength of this interaction is characterized by the gas parameter, $r_s$ (in units of the effective Bohr radius $a_B^*$), which is related to the average density of electrons, $\bar{n} \equiv \int n^2({\bf r})d{\bf r}/N$, by $r_s \equiv (\pi \bar{n})^{-1/2}$ (in 2D). More physically, $r_s$ is essentially the ratio between the potential and kinetic energy of the system, justifying its identification as the interaction strength. We tune $r_s$ by varying $\omega$ in $V_{\rm con}({\bf r})$; this makes the confining potential more narrow or more shallow, making the average density at fixed $N$ larger or smaller, thus controlling $r_s$.

The confining potential prevents the electrons from flying apart from each other, and thus an extra positive background charge is unnecessary for the stability of the dot. In the limit of weak interaction, the density and number of electrons in the dot are related to the strength of the confinement by
$\omega \!=\! e^2/(\epsilon m^* r_s^3 \sqrt{N})$.
In general, the electron-electron interactions tend to expand the dot, making the effective $\omega$ smaller than the bare $\omega$. This is quite significant for large $r_s$, and so the above simple relation between $\omega$ and $r_s$ breaks down.

Circular parabolic confinement is a good description for the experimental vertical dots, as well as for the few electron lateral dots where electrons sit in the central region far from the confining gates \cite{KouTaruchaRPP01}.
In this paper, we assume that the circular geometry is preserved even at large $r_s$, leaving the issue of irregular dots \cite{GhosalUJUB05} for future study.

Throughout this paper we use effective atomic units in which the length unit $a_B^*$ is $\epsilon/m^*$ times the Bohr radius $a_B$, and the energy is given in effective Hartrees, $H^* = m^*/\epsilon^2$ Hartrees. For GaAs, for example, $m^*=0.067m_e$ and $\epsilon=12.4$, leading to $a_B^*=98$\,\AA\ and $H^* = 11.9$\,meV.

We have studied the Hamiltonian (\ref{eq:Ham}) for a wide range of parameters. We varied $\omega$ between $3.0$ and $0.0075$ for $2 \!\leq\! N \leq 20$. For the range of $\omega$ considered, $r_s$ lies in the range $0.4$ to $18$.

\section{Method}\label{sec:meth}

Circular parabolic dots with noninteracting electrons are described
by single-particle orbitals, called Fock-Darwin (FD) orbitals, specified
by principal quantum number $n$ and azimuthal quantum number $l$. The energy
of the orbitals is
\begin{equation}
E_{n,l}=(2n +|l|+1)\hbar\omega.
\label{eq:FDen}
\end{equation}
The effective interaction between electrons can be built into the single particle problem within the framework of a mean field theory, such as the DFT or UHF methods mentioned in the introduction.
These introduce some interaction effects into the single-particle wave functions -- for instance, the resulting orbitals are more extended than the corresponding noninteracting Fock-Darwin orbitals.

As a starting point of our calculation, we use Kohn-Sham (KS) orbitals $\phi_{\alpha}({\bf r})$, from a DFT calculation within the local density approximation (LDA). We then construct configuration state functions (CSFs) that are eigenstates of $\hat{L}, \hat{S^2}$, and $\hat{S_z}$, each of which is a sum of a product of up- and down-spin Slater determinants,
\begin{equation}
\Phi_i^{L,S}({\bf R}) = \sum_{j=1}^m \beta_{ij} D_j^{\uparrow} D_j^{\downarrow} \;.
\label {eq:SD}
\end{equation}
${\bf R} \equiv \{ {\bf r}_1, {\bf r}_2, ..., {\bf r}_N\}$ denotes collective coordinates of the $N$ electrons, and $D^{\uparrow}$, $D^{\downarrow}$ are Slater determinants of spin-up and spin-down KS orbitals, respectively. The coefficients $\beta_{ij}$ are determined by the requirement that $\Phi^{L,S}({\bf R})$ is an eigenstate of $\hat{L}, \hat{S^2}$, and $\hat{S_z}$.

The many-body trial wave function that we use is
\begin{equation}
\Psi_T^{L,S}({\bf R}) = J \sum_{i=1}^{N_{\rm CSF}} \; c_i \Phi_i^{L,S}({\bf R})
\label {eq:psiT}
\end{equation}
where the Jastrow factor $J \!=\! J_{\rm en}J_{\rm ee}J_{\rm een}$ is the product of electron-nucleus, electron-electron and electron-electron-nucleus factors. By ``nucleus'' here we mean the center of the harmonic external potential. The detailed form of the Jastrow factor can be found in Ref.\,\onlinecite{GucluJeonUmrigarJain05}. The independent parameters to be optimized are the $c_i$ and the parameters in $J$. The linear combination of CSFs builds in the near-degeneracy correlation in the wave function, whereas the Jastrow factor efficiently describes the dynamic correlation that would otherwise require a very large number of CSFs. The Jastrow factor is so effective that only a small number of CSFs is needed.

In the very weakly interacting limit, the ground state consists of simply filling the lowest energy single-particle orbitals. Both the circular and harmonic nature of the external potential cause degeneracies in the noninteracting spectrum Eq. (\ref{eq:FDen}) and hence in the noninteracting many-body spectrum as well. Thus there is a definite ``shell structure'' in the energies of the many-particle states, much as in atoms. The shells are full for $N \!=\! 2$, $6$, $12$, $20, \ldots$; for these values of $N$ the ground state clearly has $L \!=\! 0$ and $S \!=\! 0$. For intermediate values of $N$, the orbitals are filled so that the interaction effects yield the lowest energy. For the total spin, Hund's first rule, familiar from atomic physics, applies here as well: the electrons in the open shell arrange so as to have maximum possible spin in order to gain exchange energy. For instance, the ground state for $N \!=\! 9$ has $S \!=\! 3/2$.

For all of the results shown in this paper, we use only CSFs constructed from the lowest energy shells consistent with the desired $L$ and $S$: no inter-shell excitations are used. Thus for the ground state, the CSFs we include are those for the (possibly degenerate) noninteracting ground state. For a closed shell and $L\!=\!S\!=0$, for instance, there is only one such CSF. For an open shell, there can be more than one CSF meeting our criteria; as an illustration, consider the $L\!=\!S\!=0$ state of $N\!=\!8$. Six electrons fill the lowest two shells while two are distributed among the third shell's three levels, $(n\!=\!1, l\!=\!0)$ and $(n\!=\!0, l\!=\!\pm\!2)$. The desired state can be made in two ways, by either putting both electrons in the former level or putting one in each of the latter (in a singlet state). We always include both of these CSFs in our calculations for this state. For a given $N$, we consider all $L$ and $S$ which can be obtained with CSFs built from orbitals in the lowest energy shell.

Selected cases are further checked by including more CSFs. For $N\!=\! 3$, $6$, $7$, $9$, and $20$, some $(L,S)$ values were studied with CSFs which included up to two inter-shell excitations. For the range of $r_s$ studied here, the energies of these states were not significantly changed by including these additional CSFs, giving confidence in the accuracy of our results.

We perform both variational Monte Carlo (VMC) and diffusion Monte Carlo (DMC) calculations. Both the VMC and the DMC energies are upper bounds to the true ground state energy, $E_{\rm GS}$. The VMC energy,
$E_{\rm VMC} \geq E_{\rm GS}$, is
\begin{eqnarray}
E_{\rm VMC} &=& \frac{\int \Psi_T^*({\bf R}) \,{\cal H}\, \Psi_T({\bf R}) \,d{\bf R}}
{\int \Psi_T^*({\bf R}) \, \Psi_T({\bf R}) \,d{\bf R}} \\
&=& \int d{\bf R} \, E_L({\bf R}) \, {\cal P}({\bf R})
\approx \sum_{i=1}^{N_{\rm MC}} E_L({\bf R_i}) \;.\nonumber
\label {eq:Evmc}
\end{eqnarray}
Here, (a) the ${N_{\rm MC}}$ Monte Carlo configurations are sampled from
${\cal P}({\bf R}) \!=\! |\Psi_T({\bf R})|^2/ \int |\Psi_T({\bf R})|^2 d{\bf R}$, and (b) $E_L({\bf R}) \!=\! \Psi_T^{-1}({\bf R}){\cal H} \Psi_T({\bf R})$ is the {\it local energy} which is constant and equal to the true energy in the limit that $\Psi_T$ is an exact eigenstate.
The variational parameters were optimized using the variance-minimization method~\cite{UWW88}, and some of the results were checked by using two recently developed energy minimization methods~\cite{UmrigarFilippi05,Umrigar07}.

We use diffusion Monte Carlo (DMC) to project out the best estimate of $E_{GS}$ starting with the optimized $\Psi_T$. The DMC method employs the importance-sampled Green function,
\begin{equation}
G({\bf R}',{\bf R},\tau)=\Psi_T({\bf R}') \langle {\bf R}'|{\rm exp}(-{\cal H}
\tau)|{\bf R} \rangle/\Psi_T({\bf R}) \;,
\label {eq:GF}
\end{equation}
to project out $\Psi_0({\bf R})\Psi_T({\bf R})$, where $\Psi_0({\bf R})$ is the lowest energy state that has the same spatial and spin symmetry as $\Psi_T({\bf R})$. As $\tau$ becomes large, the amplitudes of higher energy states decay exponentially compared to that of the ground state. However, $G({\bf R}',{\bf R},\tau)$ is not known exactly for the Hamiltonian ${\cal H}$ of Eq.~(\ref{eq:Ham}), and a short time approximation to ${\rm exp}(-{\cal H}\tau)$ using the Trotter formula is used repeatedly to achieve the desired projection. We use a refined algorithm following Ref.\,\onlinecite{UmrigarNightingaleRunge93} which has a very small time step error.

The mixed estimator for the DMC energy,
\begin{eqnarray}
E_{\rm DMC} = {\int \Psi_0^*({\bf R}) \,{\cal H}\, \Psi_T({\bf R}) \,d{\bf R} \over \int \Psi_0^*({\bf R}) \, \Psi_T({\bf R}) \,d{\bf R}} \;,
\end{eqnarray}
equals the ground state energy $E_0$. However, mixed estimators $O_{\rm DMC}$ of operators that do not commute with the Hamiltonian, such as the density, have errors that are linear in the error in $\Psi_T$. On the other hand, the extrapolated estimators $2O_{\rm DMC}\!-\!O_{\rm VMC}$ and $O_{\rm DMC}^2/O_{\rm VMC}$ have errors that are quadratic in the error in $\Psi_T$. All of the data shown in this paper for such operators is based on the extrapolated estimators, labeled ``QMC'', unless explicitly marked ``VMC''.

The function which minimizes the energy in the absence of any constraints has Bosonic symmetry, but we are interested in the lowest Fermionic state. Consequently, compared to the Fermionic state of interest, the Bosonic component grows exponentially fast. In the absence of statistical noise, one could still employ symmetry to obtain the Fermionic energy, but in a MC method the estimate for the energy has a statistical error that grows exponentially with $\tau$. The {\it fixed-node} approximation prevents this catastrophe by imposing the constraint that the nodes of $\Psi_0({\bf R})$ are the same as those of $\Psi_T({\bf R})$. The result of this constraint is that the fixed-node DMC energy is an upper bound to the true energy. For flexible wave functions with well-optimized parameters the fixed-node error is typically very small.

In our circular quantum dots, since states with angular momentum $L$ are degenerate with those of angular momentum $-L$ we are free to construct real wave functions by choosing the linear combinations,
$\Psi_T^{L,S}+\Psi_T^{-L,S}$ and $\Psi_T^{L,S}-\Psi_T^{-L,S}$. In this case we can use the fixed-node approximation.  If instead we choose to employ states with definite $L$ then the wave functions are complex, and we must use the fixed-phase approximation~\cite{OrtizMartinCeperley93,Bolton95}, the generalization of the fixed-node method to complex wave functions.  Usually the fixed-phase error is comparable to and slightly larger than the fixed-node error.

Technical considerations limit the present study to approximately $r_s \!\lesssim\! 18$. For larger $r_s$, one expects that more CSFs should be included in the trial wavefunction because excitations across the shell gaps produced by the interactions become more important. For example, for the $N\!=\!6$ ground state ($L\!=\!S\!=\!0$) with $\omega \!=\! 0.01$, inclusion of CSFs corresponding to two inter-shell excitations lowers the energy from $0.689228(5)$ to $0.689202(5)$. However, the variance optimization used for most of our results fails to lower $E_{\rm VMC}$ or $E_{\rm DMC}$ if there are more than $3$-$4$ CSFs in $\Psi_T$ (though it does lower the fluctuations of the energy). For a few cases, including moderate and large $r_s$ and several $N$, we have done preliminary calculations with higher orbitals by including all determinants involving promotion of two electrons across a shell gap (e.g. 10 CSFs for $N \!=\! 20$), using two recently developed energy optimization procedures \cite{UmrigarFilippi05,Umrigar07}. This, then, allows for a change in the nodes of $\Psi_T({\bf R})$. We find that typically a multi-CSF calculation produces a slight decrease in the energy for $r_s \ge 15$, with larger decreases per electron for smaller $N$. The change in the spatial structure, as in the density and pair-density discussed below, is even smaller than that in the energy.

\section{Results}\label{sec:res}

Examples of DMC energies from our calculations are presented in Table 1. For the purpose of comparison, we also include energies obtained using other techniques. The DMC energies, that are an upper bound on the true ground state energies, are lower than those obtained from the other methods, showing the accuracy of the method.

\begin{table}[t]
\caption{The ground state energy of a circular 2D quantum dot obtained by three different computational methods: diffusion quantum Monte Carlo (DMC), full configuration interaction (CI), and path-integral quantum Monte Carlo (PIMC). $N$, $L$, and $S$ specify the number of electrons in the dot, their angular momentum, and their spin. The energy is given in units of $\hbar \omega$, the characteristic energy of the external parabolic confining potential. $\lambda \!=\! 1/\sqrt{\omega}$ (in atomic units) characterizes the strength of the interactions.\\
}
\begin{tabular}{ccc|cccc|cc}
$N$  & $\lambda$  & $r_s$ &  ~$L$  &  $S$  &  ~DMC:  &  ~CI: &   ~$S_z$   &  ~PIMC:  \\
  &   & (approx) &   &    &  this work  &  Ref.\,\onlinecite{Rontani06}  &      &  Ref.\,\onlinecite{Egger99}    \\
\hline
6  &     8 & 12.5 &  0    &   0   &  60.3251(3)    &   60.64~~ &     &   \\
   &       &      &  1    &   1   &  60.4027(3)    &   60.71~~ &  1  &  60.37(2) \\
   &       &      &  0    &   2   &  60.3520(2)    &   60.73~~ &     &  \\
   &       &      &  0    &   3   &  60.3924(2)    &   60.80~~ & 3   &  60.42(2) \\
   &    10 & 16.3 &  0    &   0   &  68.9202(5)    &   69.74~~ &     &    \\
   &       &      &  1    &   1   &  69.0568(7)    &           &     &    \\
   &       &      &  0    &   2   &  68.9254(6)    &   69.81~~ &     &    \\
   &       &      &  0    &   3   &  68.9458(4)    &   69.86~~ &     &    \\
\hline
7  &     4 &  5.2 &  0    &  1/2  &  53.7351(3)    &   54.69~~ &     &    \\
   &       &      &  2    &  1/2  &  53.7265(2)    &   54.68~~ &     &    \\
   &       &      &  1    &  3/2  &  53.8183(2)    &   54.78~~ &     &    \\
   &       &      &  0    &  5/2  &  53.8357(2)    &   54.93~~ &     &    \\
   &       &      &  3    &  7/2  &  54.1633(1)    &   55.20~~ &     &    \\
   &     8 & 12.4 &  0    &  1/2  &  80.3846(4)    &           &     &    \\
   &       &      &  2    &  1/2  &  80.4925(4)    &           &     &    \\
   &       &      &  1    &  3/2  &  80.4795(4)    &           &     &    \\
   &       &      &  0    &  5/2  &  80.4135(3)    &           & 5/2 &  80.45(4) \\
   &       &      &  3    &  7/2  &  80.5146(2)    &           & 7/2 &  80.59(4) \\
\hline
8  &     2 &  2.1 &  0    &   0   &  46.8070(4)    &          &      &            \\
   &       &      &  2    &   0   &  46.8746(4)    &          &      &            \\
   &       &      &  4    &   0   &  46.7793(3)    &          &      &            \\
   &       &      &  0    &   1   &  46.6787(3)    &          &  1   &  46.5(2)   \\
   &       &      &  2    &   1   &  46.7560(4)    &          &      &            \\
   &       &      &  1    &   2   &  46.9170(4)    &          &  2   &  46.9(3)   \\
   &       &      &  0    &   2   &  47.4058(4)    &          &      &            \\
   &       &      &  3    &   3   &  47.4035(3)    &          &  3   &  47.4(3)   \\
   &       &      &  0    &   4   &  48.1810(4)    &          &  4   &  48.3(2)   \\
   &     8 & 12.2 &  0    &   0   & 102.9402(4)~~  &          &      &            \\
   &       &      &  2    &   0   & 102.9464(4)~~  &          &      &            \\
   &       &      &  4    &   0   & 103.0465(4)~~  &          &      &            \\
   &       &      &  0    &   1   & 102.9263(4)~~  &          &      &            \\
   &       &      &  2    &   1   & 102.9198(4)~~  &          &      &            \\
   &       &      &  1    &   2   & 102.9280(4)~~  &          &  2   & 103.08(4)  \\
   &       &      &  0    &   2   & 103.1965(4)~~  &          &      &            \\
   &       &      &  3    &   3   & 103.0185(3)~~  &          &  3   & 103.19(4)  \\
   &       &      &  0    &   4   & 103.0464(4)~~  &          &  4   & 103.26(5)  \\
\hline
\end{tabular}
\end{table}

Further results are organized in six topical areas: the electron density, the pair density, real vs. complex wave function, the addition energy of the dot, the ordering of the different $(L,S)$ states in energy, and the nature of the spin correlations.

\subsection{Spatial Density Profile}\label{subsec:den}

\begin{figure}[t]
\includegraphics[width=3.3in,clip]{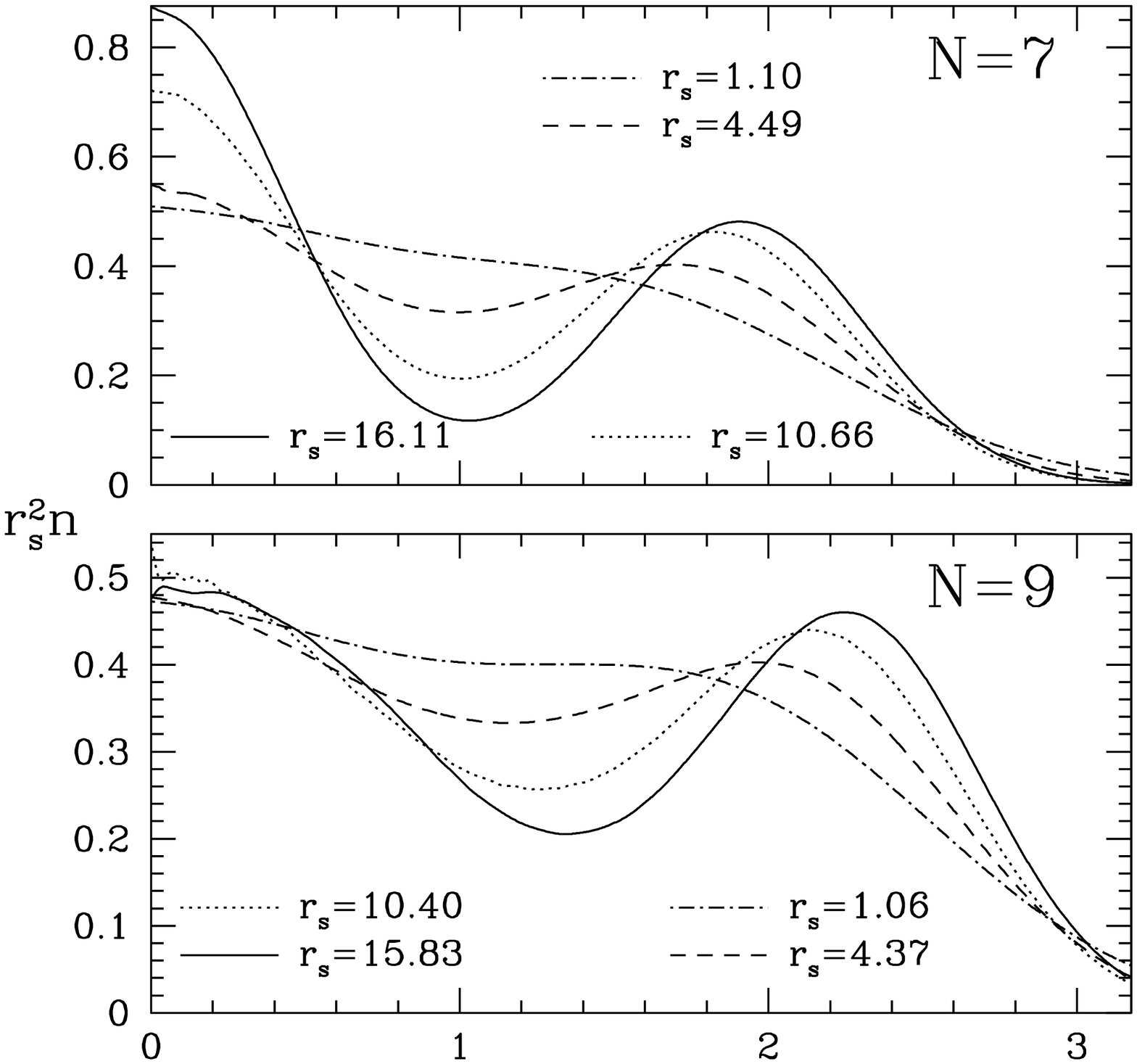}
\includegraphics[width=3.375in,clip]{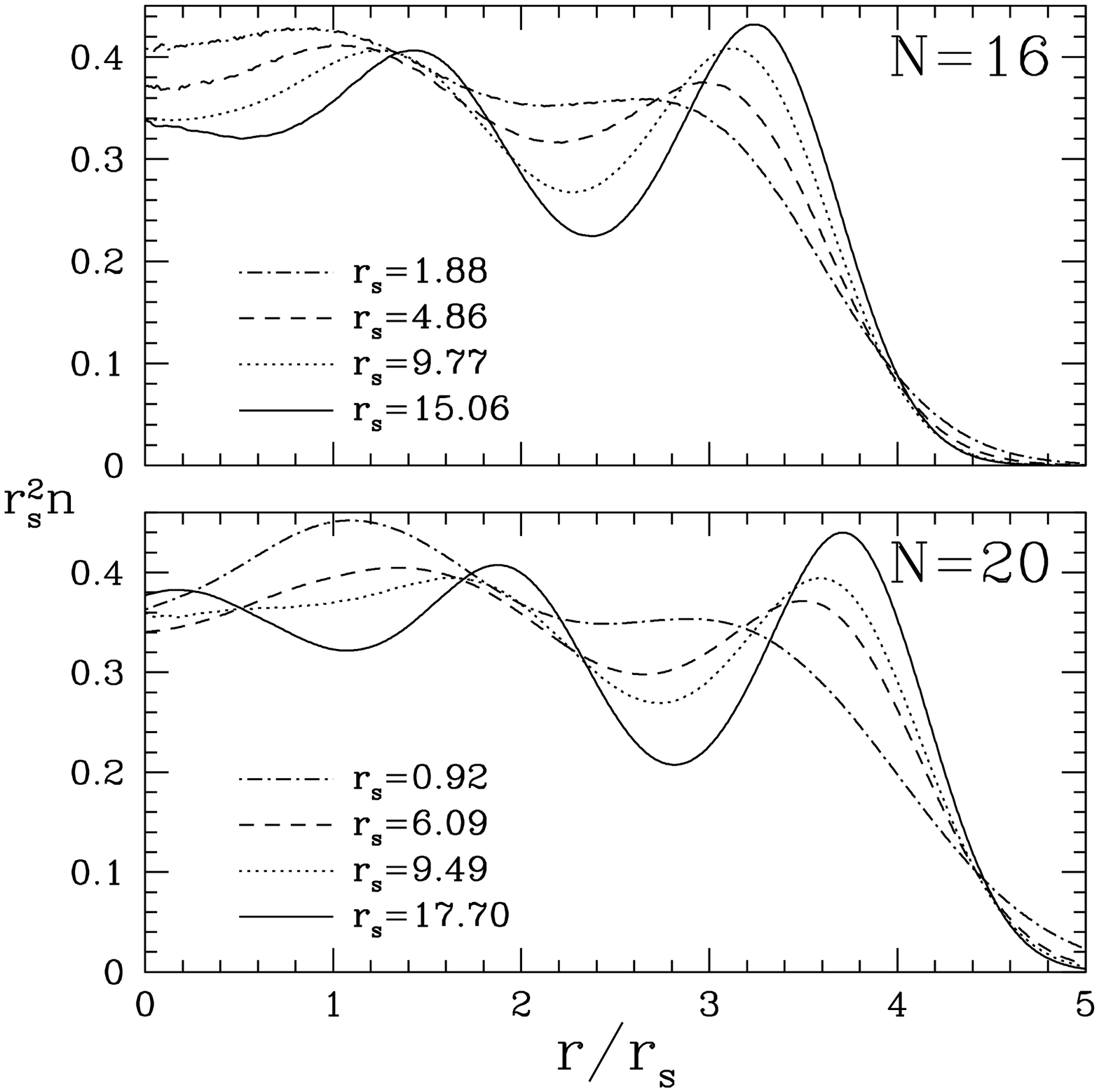}
\caption{
Radial electron density $n(r) \!=\! n_{\uparrow}(r) \!+\! n_{\downarrow}(r)$ in the ground state for 4 values of $N$ and four interaction strengths $r_s$.
\textbf{(a)}\,$N \!=\! 7$, $(L,S) \!=\! (0,1/2)$,
\textbf{(b)}\,$N \!=\! 9$, $(L,S) \!=\! (0,3/2)$,
\textbf{(c)}\,$N \!=\! 16$, $(L,S) \!=\! (0,2)$, and
\textbf{(d)}\,$N \!=\! 20$, $(L,S) \!=\! (0,0)$.
As $r_s$ increases, strong radial modulation develops leading to ring structure in $n(r)$. The number of rings for $r_s \geq 10$ is  the same as in the classical limit ($r_s \rightarrow \infty$): $2$ for the first two panels and $3$ for the last two.
[The values of $\omega$ used are:
for $N\!=\!7$ and $9$, $\omega\!=\!0.8$, $0.08$, $0.02$, and $0.01$;
for $N\!=\!16$, $\omega\!=\!0.269$, $0.06$, $0.02$, and $0.01$; and
for $N\!=\!20$, $\omega\!=\!0.8$, $0.04$, $0.02$, and $0.0075$.]
}
\label{fig:dens}
\end{figure}

The evolution of the radial density of electrons, $n(r)$, as $r_s$ varies is shown in Fig. \ref{fig:dens}. The ground state density at four different $r_s$ is plotted for four values of $N$ ($7$, $9$, $16$, and $20$). With increasing $r_s$, the electron-electron repulsion expands the system spatially; note that the size of the dot scales roughly linearly with $r_s$, as expected given the relation between $r_s$ and the average density.

The figure shows that increasing $r_s$ causes the density profile to change dramatically. For small $r_s$, the density is fairly smooth, with some weak structure coming from the noninteracting shell structure. We found that the weak structure in $n(r)$ for small $r_s$ is very similar to that obtained from Fock-Darwin orbitals consistent with the shell filling. On the other hand, large $r_s$ induces strong modulation in $n(r)$, resulting in the formation of \textit{radial rings} in the density profile. The rings become sharp with increasing $r_s$. Strikingly, once $r_s$ becomes larger than $10$-$12$, the number of rings is the same as in the classical limit \cite{BedanovPeeters94} for that $N$: one ring for $N \leq 5$, two rings for $6 \leq N \leq 15$, and three rings for larger $N$'s up to $20$. Furthermore, for these larger $r_s$, the average number of electrons in the outer ring (obtained by simply integrating the density over that region) is mostly consistent with the classical value.

If one employs wave functions of definite orbital angular momentum quantum number $L$, both the total density and the spin densities must be circularly symmetric (in two dimensions). Since our VMC and DMC methods do not break this symmetry, we obtain circularly symmetric densities in all cases. However, at sufficiently large $r_s$ even a small perturbation of the circular symmetry would be sufficient to pin electrons; in that case, the density would have the sharp peaks of a Wigner localized state. In the absence of a perturbation, the signature of localization is manifest in the pair-density, as discussed in the next subsection, even when it is absent in the density. Alternatively, a symmetry broken density profile is obtained when $L$ is not fixed; this case is treated in Section IV.C.

\begin{figure}[t]
\includegraphics[width=3.2in,clip]{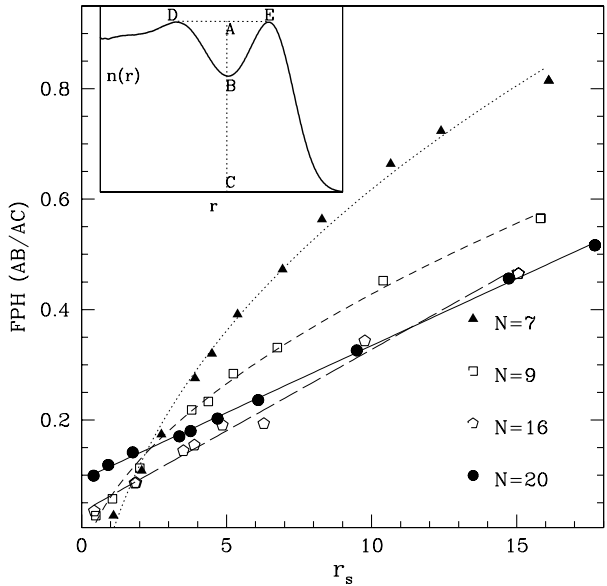}
\caption{
Evolution of the sharpness of the rings in $n(r)$ with $r_s$, which measures the radial inhomogeneity in the dot. The sharpness is quantified in terms of the ``Fractional Peak Height" (FPH), the construction for which is shown in the inset, FPH=$\overline{AB}/\overline{AC}$. The main panel shows the $r_s$-dependence of the FPH for the $N=7$, $9$, $16$, and $20$ ground states (same states as in Fig. 1). The lines show the best fit of the data to power-law behavior. The FPH grows very smoothly with $r_s$ for all $N$'s without any signature of a special threshold value.  Note that the rings sharpen faster for smaller $N$, and that the FPH is nearly linear for the larger $N$'s.
}
\label{fig:FPH}
\end{figure}

We find that the formation of the rings and the increase of their sharpness is completely \textit{smooth}. To quantify this statement, we define a quantity, the ``Fractional Peak Height" (FPH), that tracks the sharpness of the rings in $n(r)$. Its definition and dependence on $r_s$ is presented in Fig. \ref{fig:FPH}. The FPH is the ratio of the depth of the valley in $n(r)$ between the two outer rings to the `average' height of the rings. The construction is shown in the inset of Fig. \ref{fig:FPH}, where the FPH is the ratio of the two lengths $\overline{AB}/\overline{AC}$. In Fig. \ref{fig:FPH} the FPH always increases with $r_s$ as expected. However, the increase is surprisingly smooth for the whole range of $r_s$ studied. We thus infer that its increase is not associated with any special value of $r_s$ signifying a threshold or `onset'. The continuous thin lines are the best fits of our QMC results to power-law behavior. Even though the best fit curve is $\sim \sqrt{r_s}$ for small $N$ (e.g. $7$ and $9$), for larger $N=16,20$, it is nearly linear. A large FPH signifies stronger radial localization as the rings tend to decouple from each other. From Fig. \ref{fig:FPH}, we see that for the larger values of $r_s$, small $N$ dots are more strongly radially localized for a similar interaction strength.

The FPH is, by construction, some sort of `peak to valley ratio', and the construction used here is not unique. However, we have checked for a few cases that a different construction (e.g. the ratio between the height of the valley and the outer peak) leaves our conclusions unchanged.

The smooth and featureless increase of FPH with $r_s$ develops naturally from the radial structure created by noninteracting shell effects, which can be physically thought of as Friedel oscillations due to radial confinement. Our results show that these oscillations smoothly grow
into strong inhomogeneities, finally leading to radial localization. This suggests that the electron-electron interactions are acting on pre-existing oscillations, namely those caused by noninteracting interference effects, and smoothly amplifying them.

\begin{figure}[t]
\includegraphics[width=3.0in,clip]{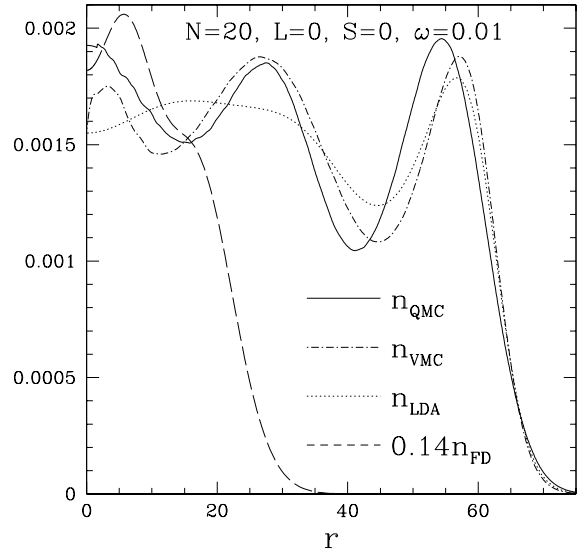}
\caption{
The effect of correlations on the radial density at different stages of approximation for the $N\!=\!20$ ground state ($L\!=\!0$, $S\!=\!0$) with $r_s \!\sim\! 15$ ($\omega\!=\!0.01$). The noninteracting density $n_{FD}$ (dashed) from Fock-Darwin orbitals is much too compact and has very weak radial structure. In the LDA result, $n_{LDA}$ (dotted), the mutual electronic repulsion makes the dot greatly expand in the radial direction compared to $n_{FD}$. Both the VMC result (dash-dotted) and QMC extrapolated estimate (solid) show much stronger inhomogeneous structure due to correlation that the QMC methods build in through the Jastrow Factor.
}
\label{fig:corrdens}
\end{figure}

\begin{figure}[t]
\includegraphics[width=3.0in,clip]{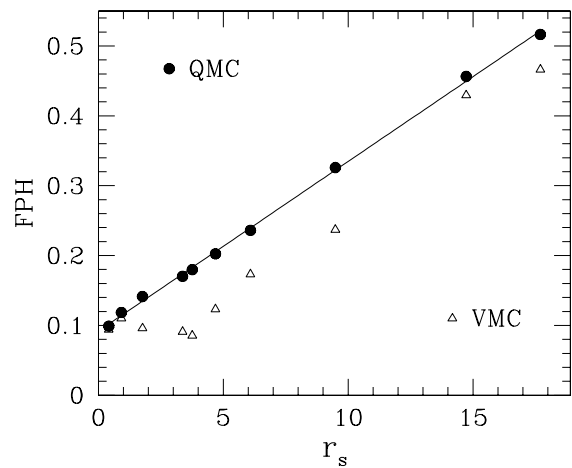}
\caption{
Growth of the fractional peak height (FPH) for the $N\!=\!20$ ground state from the QMC extrapolated estimator (solid symbols) compared to
the less accurate VMC results (open symbols). Note that the VMC estimate
alone shows a break point in the slope near $r_s \approx 4$. Thus, a DMC
calculation, with its better treatment of the interactions, is necessary to produce the smooth behavior of the FPH as a function of $r_s$.
}
\label{fig:FPHvmc}
\end{figure}

We next turn to explore the role of the electron-electron correlations in the density profile. First, we show in Fig. \ref{fig:corrdens} the density resulting from four successively better treatments: noninteracting, DFT in the LDA approximation, VMC, and finally fixed-node QMC (extrapolated estimator). The case shown is the $N\!=\! 20$ ground state with $r_s \!\approx\! 15$. As expected, we see that the interaction significantly modifies $n(r)$ at this large $r_s$, first by greatly expanding the dot at the DFT level compared to the noninteracting description, and then by amplifying the inhomogeneous structures (e.g. rings), as seen by comparing the DFT and QMC results. We note that the difference between the DFT and QMC densities is very large compared to the corresponding difference for real atoms~\cite{UmrigarGonze94}, reflecting the fact that the dot is much more strongly correlated.  Second, we compare in Fig. \ref{fig:FPHvmc} the FPH obtained from VMC to the full QMC result. Interestingly, the VMC result does show a feature at $r_s \!\approx\! 4$, namely a kink in the FPH vs. $r_s$ curve, which goes away in the full result. This demonstrates the insufficient accuracy of VMC in even the modest $r_s \!\lesssim \!5$ regime \cite{footnote3}.

\subsection{Pair-Density}\label{subsec:pden}

The issue of correlation induced localization of the individual electrons can not be addressed by looking at the density alone, which is manifestly rotationally symmetric for our Hamiltonian. The formation of radial rings in the density implies, as explained in the previous section, radial localization of electrons -- a feature special to circularly symmetric confined systems. This symmetry is, however, broken when one of the electrons is held fixed at a particular position. Other electrons then organize themselves in the dot so as to minimize their mutual interaction and kinetic energy; in particular, when the repulsion is strong enough, electrons localize at the classical positions. Therefore, to address the question of individual electron localization we turn to the pair-density.

The spin-resolved pair density, $g_{\sigma\sigma'}({\bf r}_0;{\bf r})$, is defined as the probability of finding an electron with spin $\sigma'$ at location ${\bf r}$ when an electron with spin $\sigma$ is held fixed at ${\bf r}_0$, and the total pair density is $g_T \!=\! g_{\uparrow\uparrow} \!+\! g_{\uparrow\downarrow}$. These quantities detect, in addition to radial rings, any \textit{angular} structure induced by the interactions.


\begin{figure}[b]
\includegraphics[width=3.375in,clip]{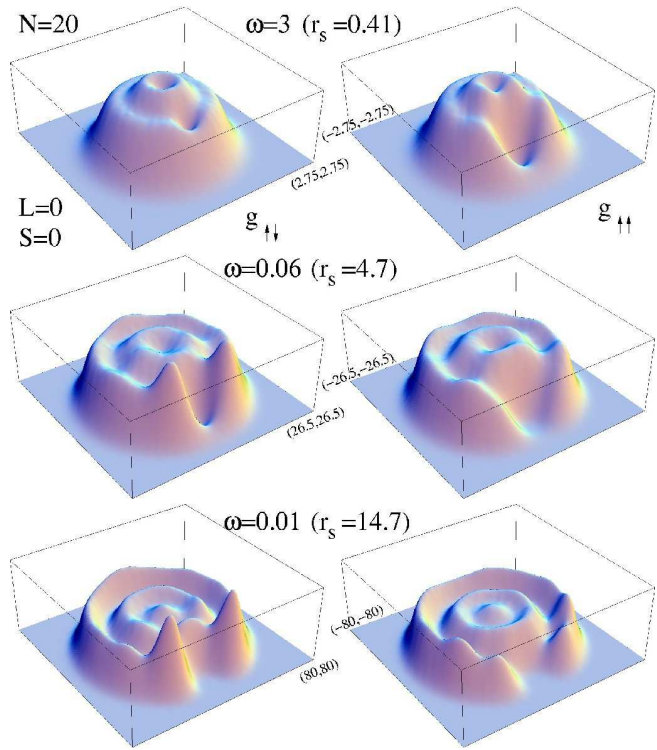}
\caption{
Evolution of the spin resolved pair-densities, $g_{\sigma\sigma'}({\bf r}_0;{\bf r})$, for the $N\!=\!20$ ground state ($L\!=\!S\!=\!0$). $r_s$ increases from the top to the bottom; $g_{\uparrow\downarrow}$ is on the left while $g_{\uparrow\uparrow}$ is on the right; ${\bf r}_0$ is chosen near the outer ring of the dot. During the initial increase of correlation until $r_s \!\sim\! 5$, the `correlation' and the `exchange' hole become fully developed. A further increase of $r_s$ produces short range order as seen from the bumps along the outer ring near ${\bf r}_0$. The range of these angular oscillations as well as their amplitude increase gradually with $r_s$, suggesting the term ``incipient Wigner localization" for this regime.
}
\label{fig:pairdens1}
\end{figure}

We present in Fig.\,\ref{fig:pairdens1} $g_{\uparrow\downarrow}$ and $g_{\uparrow\uparrow}$ for the $N\!=\!20$ ground state for weak, intermediate, and strong interaction strengths ($r_s \approx 0.4$, $5$, and $15$, respectively). The location ${\bf r}_0$ of the up-spin electron is fixed
at the outermost local maximum of the density, i.e. on the center of the outer ring. The behavior of $g_{\sigma\sigma'}({\bf r}_0;{\bf r})$ for small $r_s$ is well known: in this weakly interacting limit, the `correlation hole' (the hole around ${\bf r}_0$ in $g_{\uparrow\downarrow}$) is much smaller than the `exchange hole' (the similar hole in $g_{\uparrow\uparrow}$). This is because an opposite spin electron can come arbitrary close to the fixed one, while a same spin electron is forbidden to be in its vicinity by the Pauli exclusion principle. We see that the correlation hole gets bigger with $r_s$, becoming comparable in size to the exchange hole by intermediate interaction strength ($r_s \!\approx\! 5$).
At the same time that the correlation hole gets bigger, so do the closest peaks in the antiparallel-spin pair density but not those in the parallel-spin pair density. This is a reflection of the fact that the correlation hole integrates to $0$ while the exchange hole integrates to $-1$.

\begin{figure}[b]
\includegraphics[width=3.2in,clip]{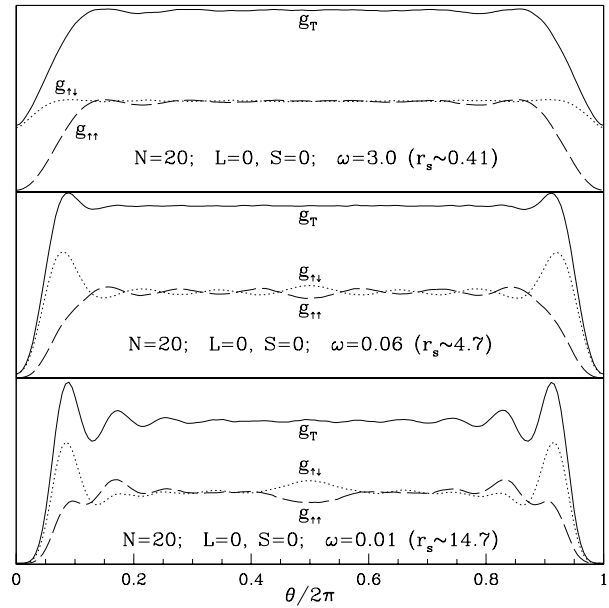}
\caption{
Evolution of the angular modulation in $g_{\sigma\sigma'}({\bf r}_0;{\bf r})$
along the outer ring for the same cases as in Fig. \ref{fig:pairdens1}.
$g_{\uparrow\downarrow}$ is shown by the dotted line and $g_{\uparrow\uparrow}$ by the dashed line. The solid line represents the spin-summed pair-density, $g_T$. The location of the fixed electron ${\bf r}_0$ defines $\theta\!=\!0$. Note the clear Friedel oscillations and the resulting spin density modulation in the middle panel. In the lower panel (strongest interactions), a strong feature at $\theta\!=\!\pi$ is present.
}
\label{fig:pairdens_angmod}
\end{figure}

Fig. \ref{fig:pairdens1} also shows that increasing $r_s$ induces weak angular modulation in $g_{\sigma\sigma'}({\bf r}_0;{\bf r})$, in addition to the radial ring structures. For small to intermediate $r_s$, this modulation can be viewed as Friedel oscillations arising due to the fixed electron; this is confirmed by noting that the period of the modulation matches the Fermi wavelength of electrons in the outer ring. Finally, we find that the angular modulation grows with $r_s$. Because of its continuous development toward sharper structure, this regime shows ``incipient Wigner localization".

To focus further on this angular modulation, we plot $g_{\sigma\sigma'} ({\bf r}_0;{\bf r})$ along the outer ring on which the up-spin is fixed in Fig.\,\ref{fig:pairdens_angmod}. We see that the angular oscillations in $g_T$, $g_{\uparrow\downarrow}$, and $g_{\uparrow\uparrow}$ are damped and weak in comparison with the radial modulation. They grow continuously as a function of  $r_s$ without any threshold value, as in the case of radial modulation. We find that the pair-density is almost featureless even for $r_s$ significantly greater than 1, while the short-range correlations set in for $r_s \geq 10$. Even at the largest $r_s$ studied here, the weakness of these oscillations suggests that the electrons remain essentially delocalized on each ring.

One of the intriguing features in the spin resolved pair-density in Fig. \ref{fig:pairdens_angmod} is a bump at $\theta\!=\!\pi$. At this position, which is diametrically opposite to the fixed up-electron, $g_{\uparrow\uparrow}$ decreases while $g_{\uparrow\downarrow}$ increases compared to their average value (the effect of the bump disappears from $g_T$). We find a similar feature for several $N$'s with different $L,S$ combinations when $r_s \geq 5$. The feature becomes more pronounced as $r_s$ increases. We lack a detailed understanding of this ``bump'' at this time; however, we suspect that its origin lies in the nature of the spin correlation between electrons, which will be discussed further in Section IV.F.

\begin{figure}[b]
\includegraphics[width=3.375in,clip]{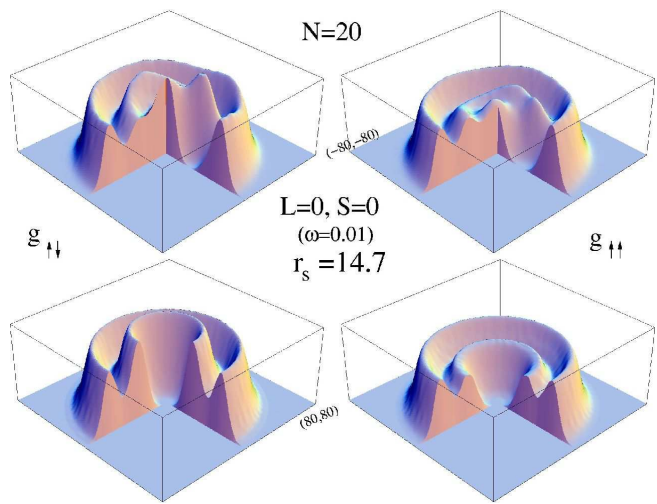}
\caption{
Spin resolved pair-densities with the fixed electron chosen on the middle ring (top panel) or in the center of the dot (bottom panel). The parameters are the same as in the bottom panel of Fig. \ref{fig:pairdens1}.
}
\label{fig:pairdens_midcent}
\end{figure}

We have also studied the pair-density with the position ${\bf r}_0$ of the fixed up-electron at different locations. Fig. \ref{fig:pairdens_midcent} shows the spin-resolved pair-densities for the $N\!=\!20$ ground state at $r_s \approx 15$ with ${\bf r}_0$ chosen on the middle ring (upper panel) and at the center \cite{footnote4} of the dot (bottom panel). These plots, together with those of Fig. \ref{fig:pairdens1}, suggest that the angular modulation is produced primarily in the same ring as the fixed electron, while the other rings remain little affected. This is an indication that even though the FPH at $r_s \approx 15$ is substantially smaller than its maximum value of $1$, the radial localization of the electrons is rather strong so that the rings essentially decouple from each other.

\begin{figure}[b]
\includegraphics[width=3.375in,clip]{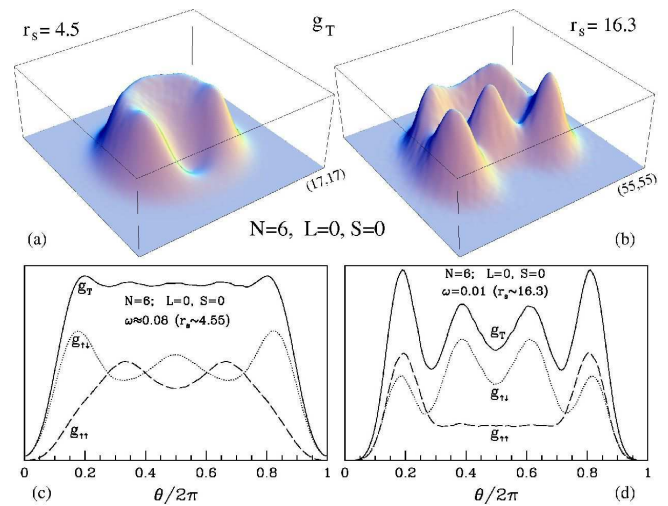}
\caption{
Pair-densities for the $N\!=\!6$ ground state ($L\!=\!S\!=\!0$, a closed shell configuration). $g_T$ for a fixed electron on the outer ring is shown for (a) $r_s \approx 4.5$ and  (b) $r_s \approx 16.3$. The corresponding angular modulation of the spin-resolved pair-densities is shown in (c) and (d). The magnitude of the angular modulation is clearly larger than that for $N\!=\!20$ at similar values of $r_s$ (compare with Fig. \ref{fig:pairdens1}): electrons are more localized in smaller dots.
}
\label{fig:Neq6}
\end{figure}

We have studied the pair-density for several different $N$ over a wide range of $r_s$. As an example, the pair-density for a small dot, $N\!=\!6$, in its ground state (a filled shell case) is shown in Fig. \ref{fig:Neq6} for three values of $r_s$. (The case $N\!=\!9$, corresponding to a half-filled shell, was shown in our previous paper, Ref.\,\onlinecite{GhosalNP06}.) In the classical limit ($r_s \!\rightarrow\! \infty$), two rings are expected for $N\!=\!6$ -- a single electron in the center and an outer ring containing the remaining five. Note that the $r_s \!\approx\! 16.3$ result is quite consistent with this classical structure. The way in which the total
modulation is shared between $g_{\uparrow\downarrow}$ and $g_{\uparrow\uparrow}$ near $\theta\!=\!\pi$ is surprising [see Fig. \ref{fig:Neq6}(d)]: all of the modulation is in $g_{\uparrow\downarrow}$ while $g_{\uparrow\uparrow}$ is constant and smaller.
We also notice that compared to $N\!=\!20$ for similar $r_s$, the angular oscillations in this case are stronger. From our extensive study, we find, using results for both pair-density and density (previous Section), that modulation caused by correlation effects is stronger when $N$ is small.

\subsection{Real vs. Complex Trial Wave-function}\label{subsec:realcmplx}

In the absence of a magnetic field, one has the choice, as discussed in Section\,\ref{sec:meth}, of working with either a complex or real wave-function. (In fact, for a circularly symmetric potential, one has that choice even in the presence of a magnetic field.)  A real $\Psi_T$ is a superposition of the degenerate states with angular quantum number $L$ and $-L$, so $\Psi_T$ is no longer an eigenfunction of $\hat{L}$. As a result, the angular part of the wave-function has $\sin(m\theta)$ and $\cos(m\theta)$ terms instead of $\exp(im\theta)$ ($m$ is an integer). While the electron density coming from a complex wave-function with definite $L$ must be circularly symmetric, a real wave-function may have angular structure for states with $L \!\neq\! 0$.

\begin{figure}[b]
\includegraphics[width=3.375in,clip]{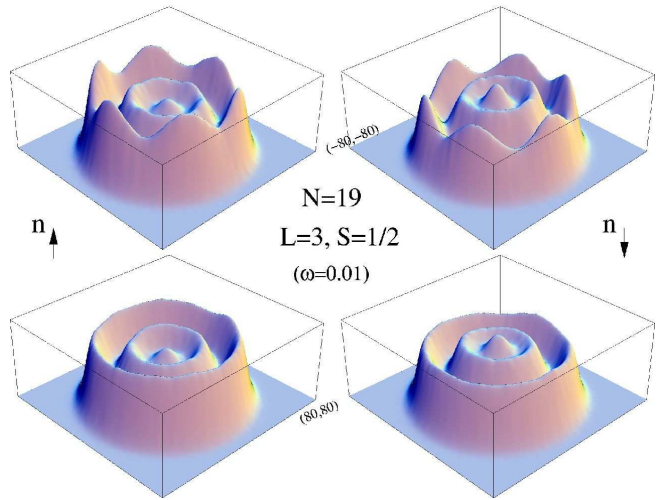}
\caption{
Spin-resolved density calculated with a real (top) or complex (bottom) wave-function.
Up- and down-spin densities for $N\!=\!19$ with $L\!=\!3$, $S\!=\!1/2$ are on the left and right, respectively. The real trial wave-function mixes angular momenta $L$ and $-L$ and so breaks the the rotational symmetry, leading to angular modulation in the density. The complex trial function has eigenvalue $L$ and shows no angular structure (within our statistics). Thus angular modulation occurs in the density if the $L$ symmetry is broken by construction; such modulation is not due to correlation effects.
}
\label{fig:complexpsi}
\end{figure}

A calculation using real wave-functions which shows angular modulation is presented in Fig. \ref{fig:complexpsi}. The spin-resolved density for $N\!=\!19$ electrons in the $L\!=\!3$, $S\!=\!1/2$ state is shown using both real and complex orbitals. We emphasize that the strong modulation seen in the top panels is \textit{not} intrinsically related to strong correlation (though we find that the modulation amplitude increases with $r_s$); in particular, the modulation is found even for small $r_s$. Furthermore, the modulation disappears in calculations using complex wave-functions, even for large $r_s$, as it must. Oscillations of a similar nature are found in the pair density -- any modulation in the density must be tracked in the pair-density. Thus we conclude that to identify the correlation induced inhomogeneities in a $L \!\neq\! 0$ state, it is important to use a complex $\Psi_T$ that has a fixed $L$.

A disadvantage of using a complex wave-function is, however, that the systematic error is typically slightly larger (fixed-phase instead of fixed-node approximation). As a result, energy estimates are more accurate when $\Psi_T$ is real.

\subsection{Addition Energy}\label{subsec:addn}

The addition energy, $\Delta^2E(N)$, is defined as the second difference of the ground-state energy with respect to the number of electrons, $N$, on the dot:
\begin{equation}
\Delta^2E(N) \equiv E_{\rm GS}(N+1) + E_{\rm GS}(N-1) - 2 E_{\rm GS}(N) \;.
\label {eq:addnen}
\end{equation}
The addition energy can be accessed experimentally through the spacing between conductance peaks in a Coulomb blockade transport measurement;\cite{MesoTran97,HeissQdotBook,MesoHouches,ReimannMannRMP02} of the various quantities that the we discuss in this paper, $\Delta^2 E$ is the simplest to measure experimentally. The leading term in the addition energy of quantum dots is the charging energy. Single particle effects and corrections to the simple charging model cause the addition energy to vary with $N$. For example, in the noninteracting limit of our Hamiltonian (\ref{eq:Ham}), the spectrum is given by Eq.~(\ref{eq:FDen}), and thus the total many-body energy is a sequence of straight line segments of increasing slope. One finds $\Delta^2E(N) \!=\! \hbar\omega$ for the ``magic numbers" $N \!=\! 2,\, 6,\, 12,\, 20,\, 30,\,\ldots$ and zero otherwise. These special $N$'s correspond to closed shell structures for which $\Delta^2E(N)$ has peaks above its baseline charging value because of the extra stability provided by the gap between shells. Weak residual interactions beyond charging produce further small peaks in $\Delta^2E$ for half-filled shells, $N \!=\! 4,\, 9,\, 16,\, 25\, \ldots$. The general occurrence of such peaks is referred to as ``mesoscopic fluctuations''; they are always present in a confined system.

\begin{figure}[b]
\includegraphics[width=3.375in,clip]{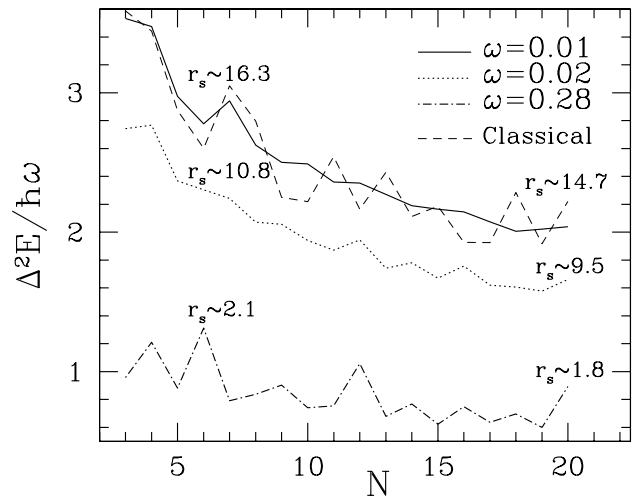}
\caption{
Addition energy (normalized by $\hbar\omega$) as a function of $N$ for three different $\omega$ and for the classical limit \cite{BedanovPeeters94} ($r_s \!\rightarrow\! \infty$). As interactions strengthen because of decreasing $\omega$, the mesoscopic fluctuations in $\Delta^2E$ become weaker. Note that this happens more readily in the small $N$ limit. Features in the $\omega\!=\!0.01$ trace at small $N$ are remarkably similar to those found in the classical limit, showing that electrons are nearly localized for small $N$.
}
\label{fig:AddEnergy}
\end{figure}

The addition energy as a function of electron number is shown in Fig.\,\ref{fig:AddEnergy} for three values of $\omega$.
As $N$ increases at fixed $\omega$, the dot grows larger but the confinement potential forces the density to increase as well, causing $r_s$ to decrease slightly with increasing $N$. In the regime of current experimental quantum dots ($\omega\!=\!0.28$ corresponding to $r_s \!\approx\! 2$), our result is similar to previous studies:\cite{ReimannMannRMP02} a base value caused by charging modulated by shell effects give strong (weak) peaks for filled (half-filled) configurations. Upon increasing the average $r_s$ ($\omega\!=\!0.04$, middle trace), we find that the strength of these peaks weakens, signaling a reduction in the meso\-scopic fluctuations as a result of electronic correlations. In particular, notice that the peak at $N\!=\!6$ has disappeared completely! The fact that the $\Delta^2 E$ curve is smoother for small $N$ indicates that correlations are stronger in that regime, a result consistent with our inference from densities and pair-densities.

Upon further decreasing $\omega$, and so increasing $r_s$, $\Delta^2E(N)$ becomes smooth at large $N$, with a weak remnant of the shell effects. However, the behavior for small $N$ has changed completely: it is no longer smooth. Note in particular a new peak that appears at $N\!=\!7$. To show the origin of this peak, we plot for comparison the addition energy in the classical limit using the ground state energy data of Bedanov and Peeters \cite{BedanovPeeters94}. Clearly, the nature of fluctuations in the classical limit is very different from those in the noninteracting limit: the peak at $N\!=\!7$ can be understood as due to the extra stability of the classical configuration. \textit{The remarkable similarity of our result for small $N$ at $r_s \approx 16$ with the classical result strongly indicates that electrons in this regime are well localized.}

We emphasize that these changes in the nature of the fluctuations of the addition energies occur gradually, and happen over different $r_s$ ranges for small and large $N$. Our study indicates that, first, the noninteracting fluctuations die out and, then, the classical fluctuations creep in.

\subsection{Reordering of States in Energy}\label{subsec:enorder}

\begin{table}[b]
\caption{The energy (in units of $\hbar\omega$) of several low-lying states (identified by $L$ and $S$) of a circular 2D quantum dot for $N\!=\!9$ and three values of $r_s$. Note the reordering of states in energy as the interaction strength increases; in particular, the $S\!=\!7/2$ state becomes nearly degenerate with the ground $S\!=\!3/2$ state at large $r_s$.
}
\begin{tabular}{cc|c|c|c}
  ~$L$  & $\;S\;$ &  $\omega=3.0$  &  $\omega=0.04$  &  $\omega=0.01$  \\
        &       & ~($r_s \approx 0.49$)~ & ~($r_s \approx 6.7$)~ & ~($r_s \approx 15.8$)~  \\
\hline
0  & 1/2 &   32.4874(1)  &   96.3487(4)  &  146.5469(9)  \\
2  & 1/2 &   32.3993(1)  &   96.3375(4)  &  146.5558(9)  \\
4  & 1/2 &   32.4387(1)  &   96.4594(4)  &  146.5746(8)  \\
0  & 3/2 &   32.3365(1)  &   96.2531(4)  &  146.4651(8)  \\
3  & 3/2 &   33.0545(1)  &   96.4594(4)  &  146.5701(9)  \\
1  & 5/2 &   34.3464(1)  &   96.6618(4)  &  146.6323(8)  \\
4  & 5/2 &   33.5959(1)  &   96.4836(3)  &  146.5892(8)  \\
0  & 7/2 &   34.7717(1)  &   96.4718(3)  &  146.4651(7)  \\
1  & 9/2 &   36.6964(2)  &   96.8211(2)  &  146.6712(7)  \\
\hline
\end{tabular}
\end{table}

The shell structure present in circular quantum dots in the weakly interacting limit suggests, in analogy with atoms, that the orbitals within a shell should be filled in the ground state so as to follow Hund's rules: the spin should be the highest possible (first rule) and the angular momentum should be the maximum consistent with the first rule (second rule). Implicit in the rules is the definition of a shell, containing orbitals that are either exactly or approximately degenerate in energy.  In atoms, orbitals with the same principal quantum number, $n$, and the same angular quantum number, $l$, are exactly degenerate, while orbitals with the same $n$ but different $l$ have an accidental degeneracy in the noninteracting limit because of the Coulomb potential. The self-consistent potential splits this degeneracy to favor states with low $l$. In quantum dots, orbitals with the same radial quantum number, $n$, and the same $|l|$ are exactly degenerate, while orbitals with the same value of $2n+|l|$ have an accidental degeneracy in the noninteracting limit because of the harmonic potential. The self-consistent potential splits this degeneracy to favor states with {\it high} $l$. In atoms, the gain in exchange energy is insufficient to overcome the energy splitting and so orbitals with the same $n$ and $l$ constitute a shell for the purpose of defining Hund's rules. In contrast, for dots, for the range of $r_s$ of interest, the exchange energy overcomes the shell splitting and so orbitals with the same $2n+|l|$ define a shell. Since the exchange interaction is present in both atoms and dots, agreement with Hund's first rule is expected for weakly interacting dots. However, there is no reason to expect Hund's second rule to hold, but one can enunciate a modified Hund's second rule -- states that occupy orbitals with high $|l|$ are favored -- which should hold for weakly interacting dots.

\begin{figure}[b]
\includegraphics[width=3.375in,clip]{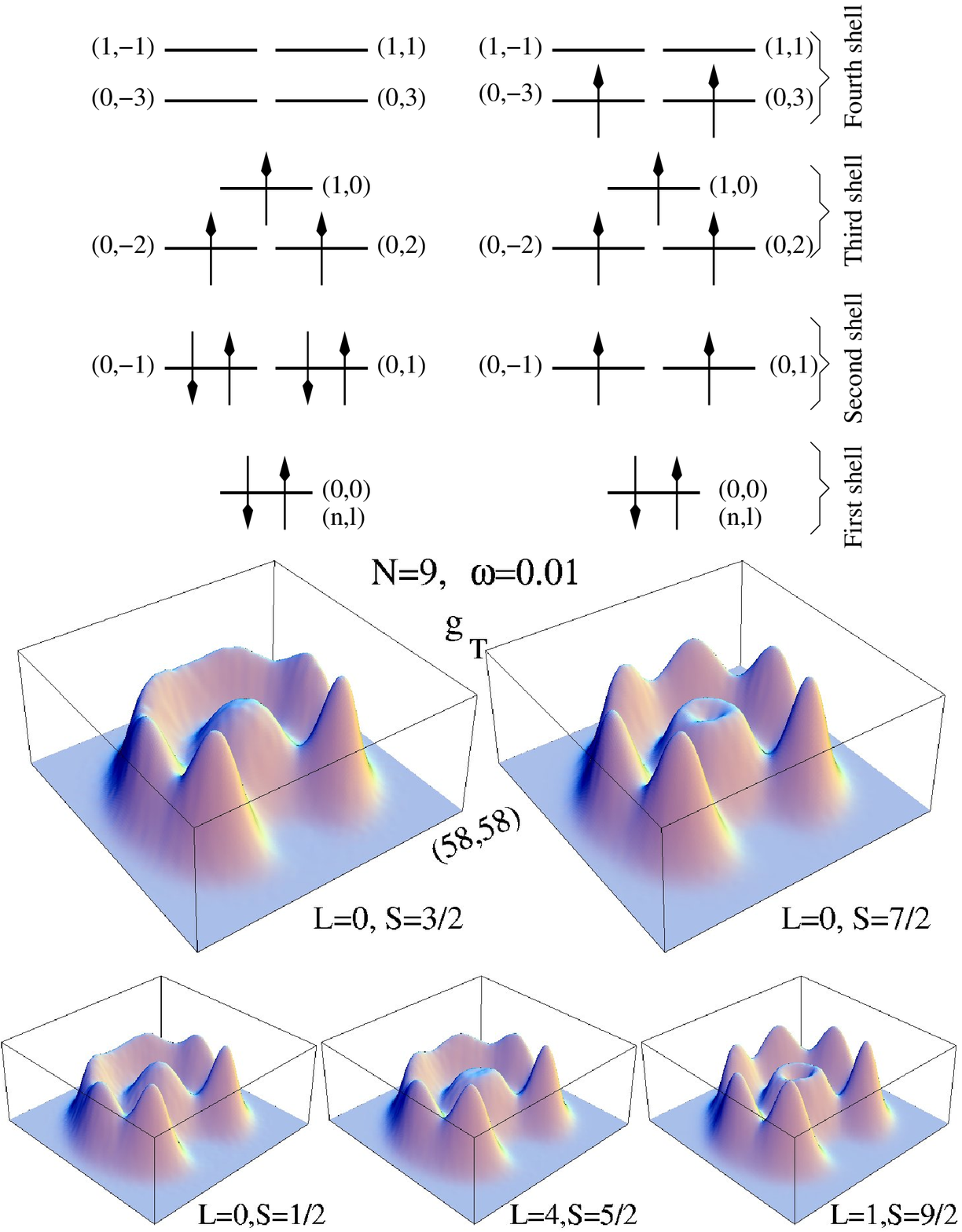}
\caption{
Pair density and level structure for $N\!=\!9$ and $\omega\!=\!0.01$ ($r_s \!\approx\! 15.8$). The spin-summed pair density, $g_T$, is shown for the lowest energy state in each of the five possible spin sectors; a spin-up electron is fixed on the outer ring of electrons. The states in the top row are the lowest in energy and degenerate within our statistical error [$(0,3/2)$ is the Hund's rule state]. For each of these states, the filling of the orbitals in the trial wavefunction is shown. The classical configuration for this $N$ is $2$ electrons in the center with $7$ in the outer ring; this configuration is clearly discernible in the pair densities of the higher spin states.
}
\label{fig:N9}
\end{figure}

Taking $N\!=\!9$ as an example, we show in Table II results for the low-lying states at three values of $r_s$; the level structure and pair densities are shown in Fig.~\ref{fig:N9}. Note that the ground state has $S\!=\!3/2$ as expected from Hund's first rule. At small $r_s$, the higher spin states lie at progressively higher energy because they involve promotion across one or more shell gaps: the kinetic energy cost of such a promotion is too large for any interaction effects to overcome. For instance, a transition from the $(L,S)\!=\!(0,3/2)$ state to the $(0,7/2)$ state involves promoting the spin-down electrons in the $(n,l)\!=\!(0,1),(0,-1)$ orbitals to spin-up electrons in the $(n,l)\!=\!(0,3),(0,-3)$ orbitals. The $S\!=\!9/2$ state requires a further promotion of the spin-down electron in $(n,l)\!=\!(0,0)$ to the $(1,1)$ orbital.


Note that $(L,S)\!=\!(0,3/2)$ is the ground state for all three values of $r_s$, in agreement
with Hund's rules.
As expected, there are numerous violations of Hund's original second rule in the different
spin cases but the modified rule discussed above holds at small $r_s$.
For example, (1) for $N\!=\!9$ in the $S\!=\!1/2$
sector (Table II), the $L\!=\!2$ state (with the $(n,l)\!=\!(0,2)$ orbital doubly occupied and
the $(n,l)\!=\!(0,-2)$ orbital singly occupied) has a lower energy than the $L\!=\!4$ state
(with the $(n,l)\!=\!(0,2)$ orbital doubly occupied and
the $(n,l)\!=\!(1,0)$ orbital singly occupied),
, and, (2) for $N\!=\!8$ (see Table I), the ground state at
$r_s \!\approx\! 2$ is $(L,S)\!=\!(0,1)$ rather than $(2,1)$.

As a function of $r_s$, there are two examples in Table II of reordering of excitations: First, in the $S\!=\!1/2$ sector, the $L\!=\!0$ and $L\!=\!4$ excited states interchange their position by $r_s \!\approx\! 6.7$, and then by $r_s \!\approx\! 16$ the $L\!=\!0$ state replaces $L\!=\!2$ as the lowest energy state. Second, as the strength of the interactions increases, the $S\!=\!7/2$ excitation becomes progressively of lower energy, interchanging with both $S\!=\!5/2$ states by $r_s \!\approx\! 6.7$ and then with the three $S\!=\!1/2$ states by $r_s \!\approx\! 16$.

In general, we find that Hund's first rule is very robust: according to our data, it can be used for the ground state spin throughout the range $4\!\le\!N\!\le\!20$ and $\omega\!>\!0.01$, corresponding to $r_s\!\lesssim\!16$. For instance, Tables I and II show that the first rule works well for $N\!=\!6$-$9$, both at small and large $r_s$. In this respect we disagree with the PIMC results of Ref.\,\onlinecite{Egger99} which predicted violation for all these cases. We believe the lack of sufficient statistical accuracy in Ref.\,\onlinecite{Egger99} led to that erroneous conclusion. Note, for instance, that the statistical error in the PIMC energies is often larger than the energy differences between different $S$ states that we find.

The only problematic case for Hund's first rule is $N\!=\!10$. Here for $\omega \!\leq\! 0.04$, we find that a state with spin $0$ [the state $(0,0)$] becomes essentially degenerate within the accuracy of our calculation with the expected $S\!=\!1$ ground state $(2,1)$. Determining the true ground state in the low-density regime in this case must await further work. The near degeneracy of these two states has been noted previously \cite{HiroseWin02,PederivaUmrigar00,GucluGuo03}. The reason is clear from the level diagram in Fig.~\ref{fig:N9}: the kinetic energy difference in moving an electron between the $n\!=\!1$ and $0$ orbitals (due to the small splitting between states having different principal quantum numbers $n$) very nearly equals the exchange energy of one pair of spins.
In fact, $N\!=\!10$ is the only $N$ in our range for which promotion of an electron across the intra-shell
gap results in a gain in the exchange energy of just one pair of electrons, explaining why it is the only inconclusive case.

The single example of a clear violation of the first rule that we have encountered is for $N\!=\!3$ when $\omega \!\leq\! 0.028$: the $(0,3/2)$ state is lower in energy than the Hund's rule ground state $(1,1/2)$. Our results for the energies are essentially the same as from a configuration interaction (CI) calculation~\cite{Rontani06} and hence not shown here. (For small $N$, CI calculations produce very precise energies.) The first rule violation in this case is probably due to the fact that instead of doubly populating the spatially localized $l\!=\!0$ orbital, electrons prefer to populate the more delocalized $l\!=\!1$ orbital.

We see a trend toward violation of the first rule at our largest $r_s$ for small $N$; this may indicate an actual violation for $r_s$ larger than was possible in this study. For example, for $N\!=\!9$ and $\omega\!=\!0.01$ in Table II, the highly spin-polarized state $(0,7/2)$ becomes degenerate with the usual Hund's rule ground state $(0,3/2)$, to within our numerical accuracy. It is clear from the shell structure why $S\!=\!7/2$ is favored: once the gain in exchange energy favors promotion of a spin-down electron in orbital $(0,1)$ across two shell gaps to a spin-up electron in $(0,3)$, the $(0,-1)$ electron will be promoted to $(0,-3)$, thus explaining why $S\!=\!7/2$ becomes lower in energy than $S\!=\!5/2$. On the other hand, $S\!=\!9/2$ requires promotion across three shell gaps and so is not necessarily favored. Similarly, it is expected that for $N\!=\!6$, the relative energy of state $(L,S)\!=\!(0,2)$ would decrease substantially at large $r_s$ compared to both the Hund's rule state $(0,0)$ and the fully polarized state $(0,3)$. For $N\!=\!7$, the corresponding state is  $(0,5/2)$. This is indeed the result seen in Table I.

Finally, we present the spin-summed pair-densities, $g_T$, for different spin states for $N\!=\!9$ in Fig. \ref{fig:N9}. It is clear that the more polarized states are better localized: as expected, exchange acts to keep the electrons apart. This is a rather generic feature in our study for a wide parameter range.


\subsection{Nature of Spin correlation}\label{subsec:spcor}

\begin{figure}[b]
\includegraphics[width=3.375in,clip]{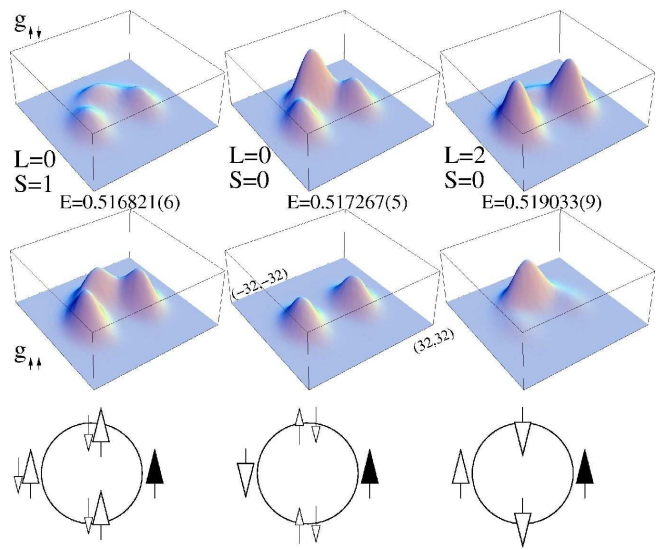}
\caption{
Nature of spin-correlation in the three low-lying states for $N=4$ at large $r_s$ ($\omega \!=\! 0.01$, $r_s \!\approx\! 18$). Top and middle panels show $g_{\uparrow\downarrow}$ and $g_{\uparrow\uparrow}$, respectively. Notice that the probability of finding electrons at a given location depends crucially on the $L$ and $S$ quantum numbers. The schematic spin-correlation is given in the lower panel. Interestingly, antiferromagnetic correlation occurs for $L\!=\!2$, $S\!=\!0$ which has the highest energy among these three states.
}
\label{fig:Neq4}
\end{figure}

The last topic we address is the nature of the spin correlation in the dots for our larger values of $r_s$. For values of $r_s$ by which angular modulation in the pair-density has developed, the radial localization is already strong, restricting the motion of the electrons in the radial direction. Thus, the electronic behavior in this limit is perhaps best described in terms of a quasi one-dimensional (1D) system on a circular ring. Effective spin interactions and the resultant correlations have been studied extensively in quasi-1D and 2D over the years \cite{Thouless65,VoelkerChakravarty01,Matveev_spinpolar06}. Recently, for instance, Matveev and collaborators have shown that in a quasi-1D system ``ring exchange" processes dominate at intermediate $r_s$, leading to novel ground states \cite{Matveev_spinpolar06}.

Our results on circularly symmetric quantum dots suggest that the the nature of the spin structure depends on the angular momentum quantum number $L$. We illustrate this by taking the simplest example, a $4$-electron dot. Fig.\,\ref{fig:Neq4} shows the spin-resolved pair-densities for three different low-lying states $(L,S) \!=\! (0,1)$, $(2,0)$, and $(0,0)$ for $r_s \!\sim\! 18$. The fixed spin-up electron is at $\theta\!=\!0$ on the single ring. We find that the probability of finding the other electrons is maximum at the classical locations--- $\theta \!=\! \pi/2$, $\pi$, and $3\pi/2$. But the spin of the electron at these locations depends crucially on the quantum numbers $L$ and $S$ of the state. The ground state is $(0,1)$ (the Hund's rule state), for which the unconstrained electrons ($2$ up and $1$ down) are equally likely to occupy any of the remaining three classical positions.

More interesting situations occur for the states $(2,0)$ and $(0,0)$, for which there are an equal number of up and down electrons and which differ only in total angular momentum. The $L\!=\!2$ state shows clear antiferromagnetic correlations. The $L\!=\!0$ state is rather unusual: the $\theta\!=\!\pi$ position is occupied by a down electron, while the remaining up and down electron are equally distributed over locations $\theta\!=\! \pi/2$ and $3\pi/2$. We interpret this equal weight of up and down electrons as representing a spin lying in the plane; thus, this state corresponds to a spin-density wave with wave-vector $\pi/2$. It is interesting to note that the $L\!=\!0$ state has lower energy than the $L\!=\!2$ state (energies are given in Fig.\,\ref{fig:Neq4}) -- the anti-ferromagnetic state is \textit{disfavored}. The ordering of these two states is opposite to that in the weak $r_s$ limit and is another example of a violation of Hund's second rule. How the value of $L$ is responsible for the spin correlation (for a given $S\!=\!0$) is not settled yet and will be pursued in future studies. Similar results were found for other cases of small $N$ as well; for large $N$, there is little angular localization of electrons on each ring, making any conclusion unreliable. However, notice that the unusual spin correlation here results in an unexpected surplus of down spins directly opposite the fixed up spin; to that extent, the feature here for $N\!=\!4$ is similar to that at $\theta\!=\!\pi$ noted for $N\!=\!20$ in Section IV.B. above.

\section{Discussion}\label{sec:discus}

The emerging picture for the correlation-induced inhomogeneities in circular dots appears to be quite different from that in the bulk. First, the absence of translational symmetry in the radial direction introduces radial localization of the electrons in rings well before individual electrons localize. This is clearly against the conventional notion of a single transition or cross-over from a weak to strong correlation regime. The radial localization can be tracked by the density due to the broken translational symmetry; angular localization is reflected in the pair-density -- a quantity that does not respect the rotational symmetry of the Hamiltonian. It is clear from our study that at a given $r_s$, radial localization is stronger than angular. This is expected due to the circular geometry. We also note here that the rings in the density could in principle be directly observed using scanning tunneling microscopy. In some of the low-density electron gas systems, resolution may be limited by the fact that the electrons are buried below the surface \cite{lowdens2DEG}. Systems have been developed, however, in which the electron gas is near or at the surface \cite{Bell97,Weisendanger00,Hirayama01a}, and it is these systems which provide the best opportunities for observation of density rings.

Second, the transition between the weak and strong correlation regimes is surprisingly broad. In fact, the completely smooth evolution of the FPH with $r_s$ suggests the absence of any cross-over scale at all. Note that the ``smoothness" goes far beyond the the usual ``rounding" of a phase transition in a finite system in which a distinct change of slope occurs in the cross-over region.

Third, the correlation strength depends not only on $r_s$ but also on the number of electrons in the dot: it is typically stronger for smaller $N$, as evident from our results for FPH, pair-density, and addition energy. In particular, the nature of the mesoscopic fluctuations in addition energy is very different for small and large $r_s$. While they are determined by the noninteracting shell-effects for small $r_s$, the structure of the classical configurations dictates the fluctuations for large $r_s$. As $r_s$ increases, first the noninteracting fluctuations are smoothed out, and then the classical fluctuations set in. Our results indicate that the value of $r_s$ where the change takes place depends on $N$. Thus for confined systems, the change from the weak to strong correlation regime is strongly non-universal.

Fourth, the ground state spin is consistent with Hund's (first) rule throughout the range of our study, $4\!\le\!N\!\le\!20$ and $r_s\!\lesssim\!18$. At large $r_s$, the excitation energy of certain strongly spin-polarized states become small but they do not become the ground state. It would be interesting to see if these polarized states become energetically favorable for even larger $r_s$. The extent of inhomogeneity in the dot shows a clear dependence on the spin state, with stronger localization occurring for larger spin polarization. While the degree of electron localization is insensitive to the angular momentum $L$ of the state (for the small $L$ here), the nature of the spin-correlation (its spatial pattern, for instance) is closely connected to $L$ in an intriguing manner.

These conclusions disagree with much of the previous literature, the majority of which predicts a single, small cross-over scale accompanied by significant deviations from Hund's rule. Because our method yields the lowest energies for quantum dots to date (the method is strictly variational) and involves less severe approximations, we believe the present results to be more accurate. Much of the problem with the earlier results stems from either associated approximations or inappropriate criteria for the cross-over.

Finally, we close with a comment on the connection between these results for circular dots and those for the bulk two-dimensional electron gas. We speculate that with increasing $N$, the deep interior of the dot will behave much like the bulk system while the radial rings will persist near the boundary, reflecting the circular confinement. Because of the rings, much of the physics discussed here will persist locally near the boundary.

\begin{acknowledgments}
We would like to thank X. Waintal for valuable conversations. This work was supported in part by the NSF (grants DMR-0506953 and DMR-0205328). AG was supported in part by funds from the David Saxon Chair at UCLA.
\end{acknowledgments}


\bibliography{qdotqmc,footnotes}

\begin{thebibliography}{71}
\bibitem[{foo({\natexlab{a}})}]{footnote3}
\bibinfo{note}{We call the reader's attention to this inadequacy of VMC mainly
  because it has been used extensively on occasion without a subsequent DMC
  calculation \cite{HarjuNieminen02}.}
\end{thebibliography}

\end{document}